# An exploration of temporal coherence of light through holography

Alexandre Escarguel, Céline Martin

Aix Marseille Univ, CNRS, PIIM, Marseille, France

Contact : alexandre.escarguel@univ-amu.fr

**Abstract** – The coherence time $\tau_c$ of light is an important physical parameter in fundamental and applied optics. Therefore, it is crucial that students understand its meaning. In this work, this notion is addressed through holography by simple experiments for bachelor and master students at Aix-Marseilles University. The coherence time of light sources used to make holograms is one of the key parameters for the success of such experiments: it must be large enough so that an optical wave train divided in two parts (reference/objects beams) interferes with itself to record the 3D shape of the object on the photosensitive medium. That is why laser sources, with much longer $\tau_c$ than other light sources, are commonly used to make holograms. We have recently worked with students on this problem: is it possible not to use a laser to make holograms? Which coherence time do we need for which hologram? After a study of the coherence time of different light sources (HeNe laser, red Light-Emitting Diode (LED) with and without 1 nm wide interference filter and high pressure mercury lamp with green filter) by emission spectroscopy and by interferometry, an extremely simple experimental setup has been used to realize holograms of a coin with them. The observation of these holograms coupled to a scan of the coin shape with an optical microscope enabled deducing information on the coherence length of these light sources.

## Introduction

Holography was invented by D. Gabor in 1948 [1], long before the invention of laser in 1960. Since the sixties, this "optical scalpel" found numerous applications in human society: imaging [2, 3], microscopy [4], displays [5], or environment [6, 7]. Recently, Kumar R. published a wide review of emerging scientific and industrial applications of digital holography [8]. From that time, it became quite easy to make a hologram because laser sources have much higher coherence than other ones. In parallel, holography with incoherent sources was also developed. Incoherent digital holography is an optical technique of 3D imaging with incoherent sources first proposed by Lohman during the sixties [9]. Tahara [10] recently published a review on this topic.

Holography is a great tool for pedagogic purposes: the realisation of a hologram is extremely motivating and enables the exploration of some important fundamental principles of optics and several applications (holographic optical elements, interferometry…). History of holography began at the end of the nineteenth century when G. Lippmann proposed a way of recording colors of photographed objects by white light interference with itself [11]. D. Gabor used the same idea but to record the relief of a scene instead of its colors. In 1962, Y. Denysiuk [12] combined these concepts to propose "Lippmann reflection holograms". For the first time, his techniques allowed the realization of holograms that could be seen with white light. However, unwanted vibrations can cause holography experiments to fail. An optimized Denysiuk holography setup was developed at Aix-Marseilles University to propose a simple setup to realize holograms in noisy environment [13, 14, 15].

Holography is based on a simple principle: interference between two coherent light beams. In figure 1, the beam from a laser is divided into two parts by a diverging lens L, and spreads out over the object O as well as on the photosensitive plate P. The 'object beam' OB is the beam obtained by the diffusion/reflection of light on the object, while the 'reference beam' RB comes directly from the laser. Interference between the two beams OB and RB occurs only if their path difference $L_{pd}$ is less than the light source coherence length $L_c$. For a typical holography setup, $L_c$ must be larger than a centimeter. Then, sunlight or non laser sources with very short coherence time (Table 1) cannot be used to realize such experiment. Only lasers with typical coherence length ranging from centimeters to several kilometers allow to realize holograms easily.

The coherence length of a light source is classically obtained using interferometric set-ups as Michelson or Mach-Zehnder interferometers. For low coherence sources, as LEDs, a simple way to measure it is the measurement of the light spectral width with a low resolution spectrometer. It is also possible to use more sophisticated techniques [16, 17].

In the frame of holography teaching at Aix-Marseilles University, students had to study the question of the value of $L_c$ needed to realize particular holograms with low path difference $L_{pd}$. In other words, is it possible to do a hologram of a simple coin, with a limited depth, without a laser? Instead of a detailed study of coherence properties of different light sources [18], we proposed an original way by coupling emission spectroscopy or Michelson interferometer with profilometry measurements. Moreover, an on-axis reflection holography setup extremely simple to install and very insensitive to vibrations was used, accessible to bachelor students.

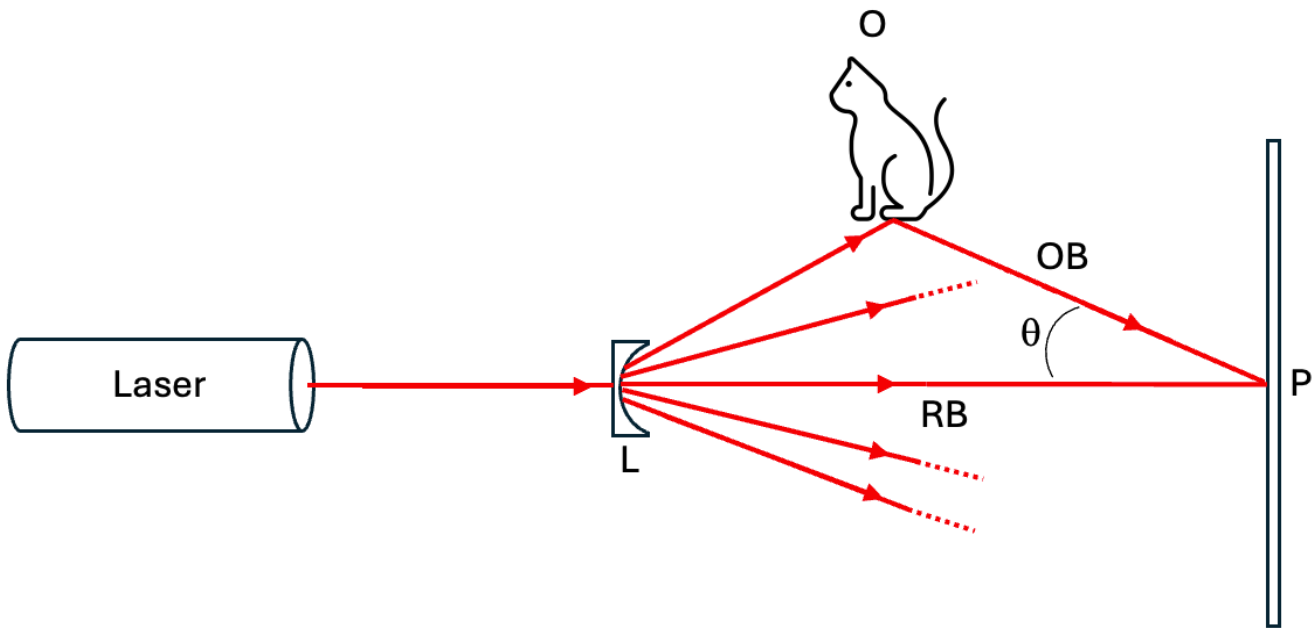

**Fig. 1**: Scheme of the hologram production principle. L: diverging lens; O: object; P: photosensitive plate; RB: reference beam; OB: object beam; and θ: mean angle between RB and OB. The lens enlarges the reference and object beams so that they cover the whole object and plate. The path difference is $L_{pd} = LOP - LP$.

This paper is organized as follows: first, the experimental setup based on previous works is presented, [13, 14, 15]. Second, coherence length of several light sources is measured by emission spectroscopy and/or Michelson interferometry in view of making hologram of a simple coin. Third, a confocal microscope is used to measure the 3D shape of the coin. Fourth, holograms of the coin are made with the light sources studied. Then, the observation of holograms coupled to profilometry results yields an estimate of the light sources coherence lengths. These estimates are compared to emission spectroscopy/Michelson interferometry results.

## 1. Denysiuk reflection holograms in noisy environment

In 2010, a holography teaching kit composed of a small case containing all the items necessary to produce holograms almost anywhere was developed at Aix-Marseilles University [13, 14, 15]. In particular, the kit allows the realization of holograms in noisy environment. Indeed, when the reference and object beams interfere onto the photosensitive emulsion, interference fringes form with details smaller than a micrometer. Then, during the realization of a hologram, a vibration of that range can prevent the hologram production. This is why classical books on holography propose to do such experiments on optical tables with vibration isolators. These precautions must be considered only after the division of the beam in two parts (reference and object beams). Then, to limit vibrations problems, the first trick is to prepare an experimental setup with that beam division as late as possible. The Denysiuk setup [12] allows this because the beam division happens only when the beams go through the photosensitive plate. Second, the use of gravity as a stabilizer permits successful holograms in noisy environment. Fig. 2 shows the periscope used in the holography teaching kit, inspired by Denysiuk's setup. A 45° mirror allows the laser beam to hit the transparent photosensitive plate from the bottom and then to be reflected/diffused back by the object which is simply put upon the plate. There is no need to physically divide the laser beam into two parts with a beam splitter and stabilize all the optical elements. In this setup, beam division only happens when the beam is crossing the plate. So, it is only necessary to have no vibrations between the plate and the object. In our case, they are stabilized with respect to each other thanks to gravity. The object is placed on the plate, which is put on the periscope. The periscope is composed of a mirror put at 45° and of three vertical rods to hold the plate. The heavier the object, the more stable it is on the plate. It is important to fix 3 spikes equally spaced under the periscope so that it holds on three points on the table for better stability. This experimental device is extremely stable, even in a non-stabilized environment with a huge number of people talking. The object beam is the one that is reflected/diffused by the object and which is coming back towards the plate, while the reference beam is the part coming directly from the laser. This produces a reflection hologram characterized by the fact that the two incident beams are from opposite sides of the plate. Note that the photosensitive emulsion must be placed on the object side to avoid internal light reflections between the two glass surfaces.

Holograms can be recorded on Ultimate silver halide plates [19]. The sensitivity is high (typically 200 µJ/cm$^2$) but a chemical processing is necessary. When using photopolymers photosensitive plates, there is no chemical development and no saturation effect. Compared to classical silver halide plates, they present the disadvantage to be at least 100 times less sensitive. However, the use of the setup shown on Fig. 2 allows successful holograms even in noisy environment. For example, using a 20 mW 532 nm DPSS laser with a 70 mm diameter diverging beam spot onto the plate, the exposure time must be at least 20 s. Fig. 2 (c) shows a resulting hologram obtained in a very noisy environment during the European researchers night in Marseilles, Oct. 2022 (https://marie-sklodowska-curie-actions.ec.europa.eu/event/2022-european-researchers-night). Successful holograms have also been made with photopolymers and a 1 mW HeNe laser. In that case, the exposure time was 120 s. An important advantage of the low sensitivity of photopolymers photosensitive plates is that absolute darkness is not necessary.

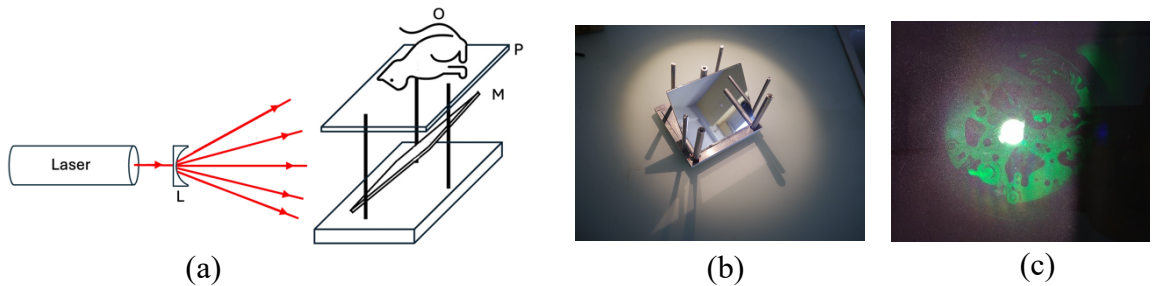

(a)                 (b)                 (c)

**Fig. 2. (a)** schematic side view of the optical setup used in the holography teaching kit: L: short focal length lens used to expand the beam; M: 45° front surface mirror; P: transparent photosensitive plate, carried by the mirror mount; O: object, carried by the plate. (b) top view of the setup. The holographic plate is put on the three thick rods. The three thinner rods are only used for an easier holographic plate positioning in darkness. **(c)** typical hologram of the back of an old alarm clock obtained with the kit in noisy environment, during the European researchers night, Marseille, Oct. 2022 (video: https://vimeo.com/859785742). Better results are obtained with metallic heavy objects. The use of photopolymers allows to avoid chemical development/processing and to work without darkness.

## 2. Measurement of coherence length of light sources

In this work, we only focus on the coherence time $\tau_c$ and the corresponding coherence length $L_c = c \times \tau_c$ ($c$=light velocity in the medium). Spatial coherence of the light source is supposed to be optimal. This means that the source diameter seen from the recording location at a distance D is small enough for interferences to occur [20]. The realization of holograms needs light sources with good coherence time [21]. $L_c$ is the mean length of light wavefront with a phase reference. It is a crucial parameter for successful holography: it must be large enough for light to interfere with itself to record the 3D shape of the object on the photosensitive medium.

Here are the questions for students: is it possible not to use a laser to make a hologram? Which coherence time for which hologram? They first studied the coherence time of several light sources (Table 1) by emission spectroscopy when possible, or with a Michelson interferometer when the spectral width is below the spectrometer resolution (1 nm). In the first case, they simply measured the full width at half maximum $\Delta\lambda$ of the light source spectrum and its central wavelength $\lambda$. Then, they calculated $L_c$ with the classical relation [21]:

$$L_c = \frac{\lambda^2}{\Delta\lambda} \qquad (1)$$

A Michelson interferometer with a movable mirror was used to estimate $L_c$ for sources with low $\Delta\lambda$. $L_c$ was directly approximated as the distance between the two extreme movable mirror positions on one arm of the interferometer for which fringes can be seen. This way of measuring $L_c$ is only an estimate because it is based on a visual determination. $\Delta\lambda$ was then deduced with relation (1). Table 1 shows the results for typical light sources. HeNe coherence length value was obtained from bibliography [22]. For the red LED with a 1 nm wide interference filter centered on $\lambda$, $L_c$ was calculated from its commercial specifications.

|  | $\lambda$ ($\mu$m) | $\Delta\lambda$ (nm) | $L_c$ (m) | Method |
|---|---|---|---|---|
| **White light desk lamp (LEDs)** | $\approx 0.5$ | $\approx 400$ | $\approx 6 \times 10^{-7}$ | Spectroscopy |
| **Red LED** | 0.6348 | 20 | $2 \times 10^{-5}$ | Spectroscopy and Michelson |
| **Red LED + interf. filter** | 0.6328 | 1 | $4 \times 10^{-4}$ | Filter specifications |
| **High pressure Hg lamp + green filter** | 0.546 | $6 \times 10^{-2}$ | $5 \times 10^{-3}$ | Michelson |
| **HeNe Laser** | 0.6328 | $10^{-3}$ | $3 \times 10^{-1}$ | Not measured |

**Table 1**: central wavelength $\lambda$, spectral width $\Delta\lambda$ and coherence length $L_c$ of typical light sources. Last column: method used. Spectroscopy measurements: $\Delta\lambda$ is measured and $L_c$ is deduced from eq. (1). Michelson interferometer measurements: $L_c$ is measured and $\Delta\lambda$ is deduced from eq. (1). For red LED, the two methods are used. They both give consistent results.

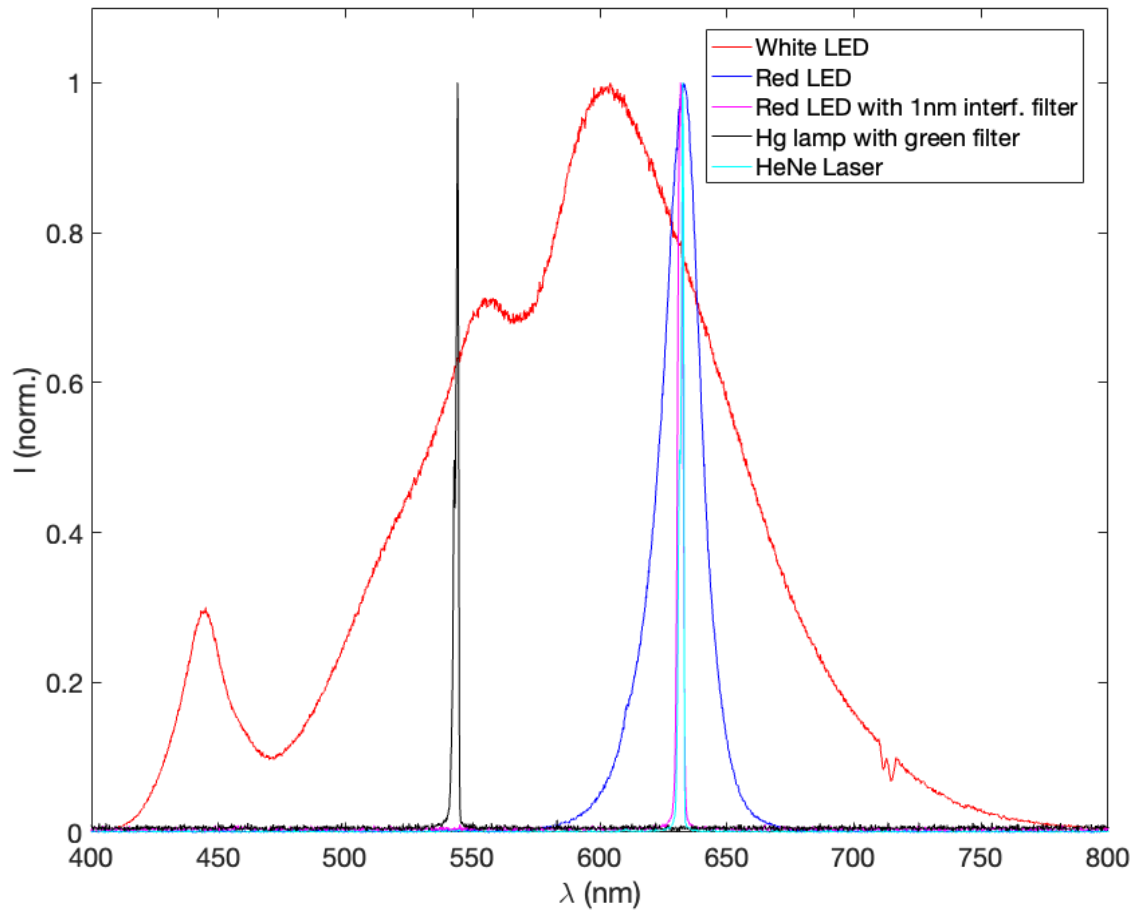

**Fig.3**: spectra of light sources presented in table 1, obtained with an emission spectrometer. The spectrometer resolution is too low to resolve the spectral width of emission lines of the red LED with a 1 nm interference filter, the mercury lamp with a green filter and the HeNe laser.

HeNe laser is a classical source for the realization of holograms, with a good coherence length around 30 cm. Then, holograms experimental setups with such light sources can have a large path difference $L_{pd}$. At the opposite, a white light desk lamp composed of LEDs has a very short coherence length, lower than 1 µm that does not allow to realize holograms. In between these two extreme cases, three sources have been analyzed: a red LED, without and with a 1 nm wide interference filter and a high pressure mercury lamp with a green filter. In the third case, the coherence length is imposed by the 1 nm wide interference filter. The used source can be any incoherent one, as long as it is small enough to have a good spatial coherence. In the third case, the green filter has a 10 nm width. It allows the selection of an intense mercury emission line at 546.08 nm whose spectral width determines the coherence length. Holograms of a simple 25 mm wide coin with a limited relief (see profilometry analysis) have been made with these three sources. The hologram has also been made with a HeNe laser, for comparison.

# 3. Profilometry for the coin relief determination

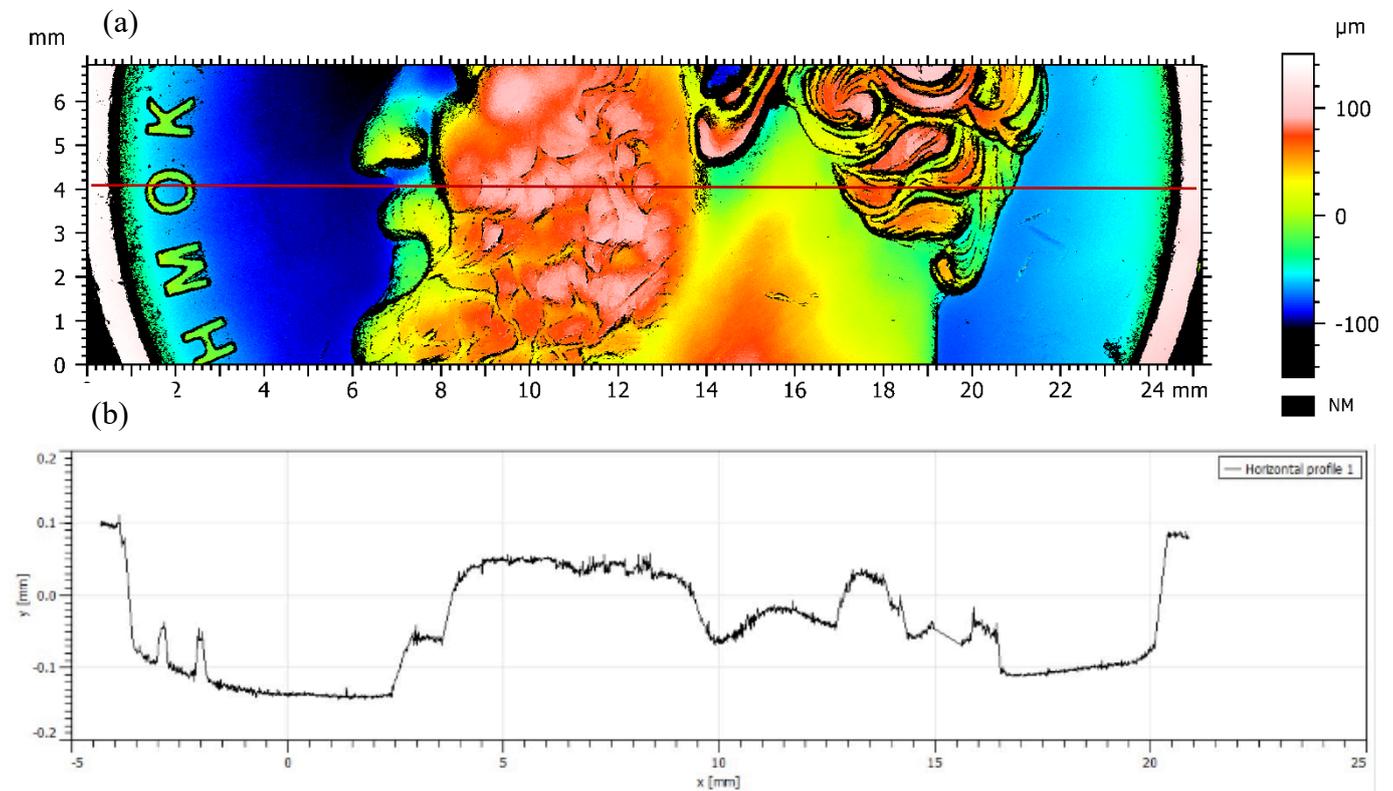

**Fig. 4**: Confocal images of the coin used to make holograms. 3D surface profilometry (a). Linear profile along the red horizontal line (b).

The 3D surface relief of the coin was measured using a confocal microscope (S neox, SENSOFAR). A light beam is focused on the sample through the objective of a microscope restricting the lateral probed area and the light reflected from the sample is recorded through a blocking pinhole in the conjugate plane (confocal aperture) to eliminate any out-of-focus light, allowing the depth of field to be reduced. A series of two-dimensional images (thin optical section) at different focal planes are acquired by a z-scanning of the objective. The laser source technology based on a microdisplay allows a fast scanning of the imaged surface without moving parts in the microscope. The images were produced using a green LED light source (530 nm), the lateral optical resolution was 0.93 µm. The 3D surface topography image (fig. 4 (a)) with a vertical resolution >80 nm is reconstructed from the series of two-dimensional images by numerical processing to deduce the height of each point by the maximum of intensity determination. At the end, a profile is extracted from the 3D surface topography (fig. 4 (b)).

# 3. Holograms with different light sources and coherence length estimations

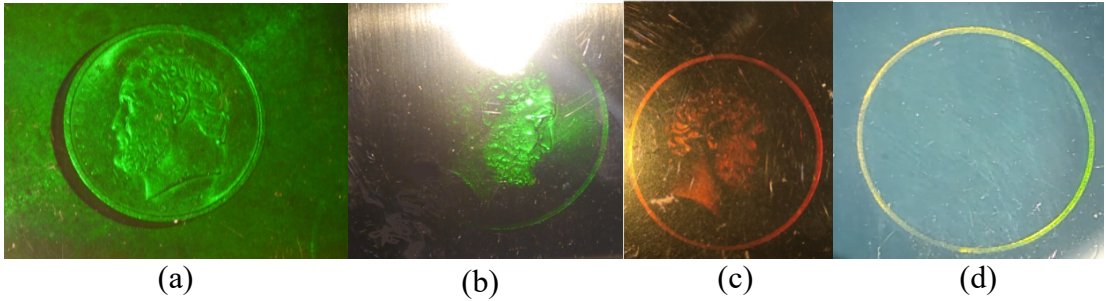

(a)                    (b)                    (c)                    (d)

**Fig. 5**: holograms of a coin obtained with a HeNe laser (a) (https://vimeo.com/961106785), a high pressure Hg lamp + green filter (b) (https://vimeo.com/915217940), a red LED + 1nm interference filter (c) (https://vimeo.com/840366078), and a LED alone (d).

The holography teaching kit setup (part 1) was used to make ~~all~~ the holograms. Along with the mercury lamp, a simple convex lens with a short focal length was used to illuminate properly the recording area on the periscope. Since the high pressure Hg lamp is an extended source, it is necessary to focus it on a 1 mm diameter pinhole to improve its spatial coherence. In this case, a second lens placed after the pinhole allowed to have a spot diameter large enough where the coin hologram is recorded. The results are shown on Fig. 5. Photographs of in-line holograms are difficult to achieve because the light source must be on the axis of observation. To help readers, short videos of holograms are available via hypertext links. Note that the color of hologram in Fig 5(a) is green because the emulsion thickness has been reduced during the chemical development process, leading to a color shift from He- Ne wavelength (632.8 nm) to 563 nm.

**HeNe**

Because of the high $L_c$ value, the whole coin and the background behind it can be seen on the hologram.

**Hg lamp + green filter**

The 5 mm coherence length of the mercury lamp allows to capture the whole coin's relief, but not the background behind the coin.

**LED + 1 nm interference filter**

This is the most interesting case. The hologram made with a red LED and an interference filter allows to capture the portrait on the coin, but not the background of the coin surface. These observations coupled to profilometry data permit to estimate a range of values for $L_c$. The face is seen on the hologram, including the beard at 150 µm from the edge of the coin. On the other hand, the bottom of the coin, at 175 µm from the edge of the coin, does not appear on the hologram. Then, we can deduce an estimate of the range for $L_c$(LED+IF):

**300 µm < $L_c$(LED+IF) < 350 µm**

The upper value of this estimate is 13 % lower than the theoretical value of coherence length (namely 0.4 mm) (Table 1).

**LED alone**

Finally, the hologram made with a LED alone only records the closest part of the coin (circular edge) in contact with the emulsion. This means that its coherence length is lower than twice the distance $d_{min}$ of the closest parts of the coin's portrait in the center. Profilometry measurements give $d_{min} = 40$ µm. Then, we can deduce that $L_c(LED) < 80 \mu m$. This is consistent with measurements made by spectroscopy or Michelson interferometry.

## Conclusion

The question of light coherence in holography has been explored by proposing experiments for bachelor and master students at Aix-Marseilles University. The following questions have been explored by students: is it possible not to use a laser to make a hologram? Which coherence time for which hologram do you need? This was done within the framework of project types teaching, one half day a week during one semester. This allows students to have enough time to explore themselves the proposed issues. Students studied the coherence length $L_c$ of several light sources by emission spectroscopy and/or Michelson interferometry:

- high pressure mercury lamp,

- red LED with a 1 nm wide interference filter,

- red LED.

In parallel, profilometry of a simple coin was used to obtain its relief. Then, its holograms were made with the three sources and with a HeNe laser for comparison. Coupled to profilometry data, this allowed to obtain estimations of $L_c$ coherent with the measured ones. The proposed set of experiments is a good opportunity for students to explore the notion of optical coherence time by experiments with various methods. Starting from a laser hologram, they can see how the decrease of coherence length progressively reduces the possible depth of recording on the hologram. The choice of a simple coin with low relief is particularly well adapted to such experiments.

## Aknowledgements

This work received support from the French government under the France 2030 investment plan, as part of the Initiative d'Excellence d'Aix-Marseille Université - A*MIDEX ( AMX-19-IET-013 – ISFIN). Authors thanks gratefully Gregory Giacometti for profilometry measurements.